%% file: kampingv2.tex
\documentclass[runningheads]{llncs}
\usepackage{fancyhdr}
\fancypagestyle{firstpage}{
  \fancyhf{}
  \fancyfoot[L]{
    \scriptsize
    This preprint has not undergone peer review or any post-submission improvements or corrections.
    The Version of Record of this contribution is published in Recent Advances in the Message Passing Interface, and is available online at \url{https://doi.org/10.1007/978-3-032-40777-1_2}
  }
}
\makeatletter
\def\sectionmark#1{}%
\def\subsectionmark#1{}%
\makeatother
\usepackage[T1]{fontenc}
\usepackage{graphicx}
\usepackage{orcidlink}
\newcommand{\myorcid}[1]{\orcidlink{#1}}
\newcommand{\corres}[1]{\textsuperscript{(\href{mailto:#1}{\color{black}\Letter})}}
\input{preamble}

\hypersetup{
  colorlinks=true,
  allcolors=blue
}
\usepackage[misc]{ifsym}
\begin{document}
\title{Concepts in Practice: \Cpp{} MPI Bindings for the HPC Ecosystem}
\subtitle{From a Standardizable Core to a Composable Interface}
\author{
  Tim Niklas Uhl\corres{uhl@kit.edu}\myorcid{0000-0001-9295-1388} \and
  Matthias Schimek\myorcid{0009-0002-6402-9016} \and
  Daniel Brommer
}
\authorrunning{T.N. Uhl, M. Schimek, D. Brommer}
\institute{%
  Karlsruhe Institute of Technology, Karlsruhe, Germany\\
  \email{\{uhl,schimek\}@kit.edu}\\
  \email{daniel.brommer@student.kit.edu}
}
\pagestyle{fancy}
\maketitle              %
\thispagestyle{firstpage}
\begin{abstract}

  The official \Cpp{} MPI bindings were removed from the standard in 2008, leaving a gap that numerous third-party libraries have attempted to fill.
  However, existing wrappers typically cover only a limited subset of MPI or target specific use cases, falling short of a general-purpose solution.
  A recent conceptual paper~\cite{DBLP:conf/pvm/AvansCGSSSSU25} proposed general design principles for modern \Cpp{} bindings based on \Cpp{}20 concepts, without committing to a concrete interface.

  We present the first concrete realization of these principles in a layered architecture.
  At the foundation, we define a core layer:
  refined \Cpp{}20 concepts formalizing the MPI standard's notion of data buffers, automatic mapping of standard \Cpp{} constructs, non-intrusive customization points for third-party types, and concept-based wrappers for MPI procedures.
  The result is a low-level native \Cpp{} MPI interface that works directly with STL containers, is highly extensible, and lends itself to standardization.
  Built on this core, we present \kampingtwo{} --- a \Cpp{} MPI library offering the convenience and memory-safety of KaMPIng~\cite{DBLP:conf/sc/UhlSHHKSS024} with composable, pipe-based syntax inspired by \Cpp{} ranges for efficient, boilerplate-free MPI programming.
  Finally, we demonstrate the core layer's broad applicability by designing lightweight adapters for GPU and performance-portability libraries, making the HPC ecosystem a first-class citizen in MPI. Kokkos views, Thrust device vectors, and SYCL buffers can be passed directly to MPI procedures, with adapter logic remaining self-contained.

  All contributions are backed by a fully functional open-source reference implementation, demonstrating the practical viability of the proposed design.

\end{abstract}

\section{Introduction}\label{sec:introduction}
\newcommand{\myintropar}[1]{\noindent\textbf{#1}}

Since the initial proposal of the Message Passing Interface (MPI) standard by the MPI Forum in 1994, its goal has been to define an interface for message passing that (1) provides \emph{convenient bindings} for the important programming languages underlying high-performance computing (HPC), (2) ensures \emph{language independent semantics}, and (3) supports \emph{heterogeneous environments}~\cite[Sec. 1.1]{mpi50}.
While the MPI standard has traditionally focused on C and Fortran interfaces, \Cpp{} continues to play an important role in the MPI community.
In MPI 2.0, the standard introduced official \Cpp{} bindings;
however, these were deprecated and subsequently removed in MPI 3.0, as they offered only limited additional functionality over the C interface while significantly increasing maintenance complexity.

Since then, \Cpp{} has evolved substantially. With the emergence of modern \Cpp{}-based heterogeneous programming models, there is a growing need for an interface that adheres to the language's ergonomics and design principles without compromising performance or portability.
However, reconciling MPI’s low-level, pointer-based interface with modern \Cpp{} abstractions presents a key challenge:
interfaces must remain efficient and portable while supporting safer and more expressive programming models.
The MPI standard already describes communication in terms of typed sequences of elements, i.e., conceptual data buffers~\cite[Sec. 3.2.2]{mpi50}.
In the C interface, this abstraction is expressed as a triple of pointer, count, and datatype.
We argue that language bindings should reflect such abstractions in a manner consistent with the host language;
in \Cpp{}, this naturally leads to representing buffers as typed objects that model a \Cpp{} concept, providing a uniform and extensible interface for MPI operations.

While numerous third-party \Cpp{} wrappers exist, existing approaches are either too limited in scope or too opinionated in design to serve as a standardizable interface, and none provides a faithful realization of this buffer abstraction as a first-class \Cpp{} concept.
This perspective motivates revisiting MPI \Cpp{} interfaces as direct realizations of the buffer abstraction in terms of native \Cpp{} concepts.
Recently, we presented conceptual design guidelines for such an interface~\cite{DBLP:conf/pvm/AvansCGSSSSU25}, based on a systematic analysis of existing MPI \Cpp{} interfaces and prior implementation experience, but without realizing a concrete \Cpp{} language interface.
The present work builds on this foundation by providing a concrete realization of these concepts as a fully functional open-source reference implementation.\footnote{Implementations of all interfaces proposed in this paper are available at \url{https://github.com/kamping-site/kamping-v2}.}
It is organized in three layers, building on top of the MPI standard as the backbone layer.
\cref{fig:teaser} shows representative examples from each layer.

\myintropar{Core Interface (\cref{sec:core}).}
We define a minimal, extensible core \Cpp{} MPI interface centered around the concept of a \emph{data buffer}, closely mirroring MPI semantics while lifting them to idiomatic \Cpp{} abstractions.
It supports common \Cpp{} constructs such as STL containers directly, without manual extraction of pointers or sizes.
By expressing MPI semantics natively in \Cpp{}, it addresses \emph{language independent semantics} without being constrained by the existing C language interface.
Its intentionally minimal design, covering both data buffers and MPI object handles with uniform, non-intrusive customization points, ensure maximal compatibility with existing MPI code and third-party types.
By staying faithful to the MPI standard's own abstractions rather than layering additional structure on top, the design lends itself to standardization as a native \Cpp{} MPI interface.

\myintropar{\kampingtwo{} (\cref{sec:kamping-v2}).}
Built on the core interface, \kampingtwo{} addresses \emph{convenient bindings} following the core design ideas of KaMPIng~\cite{DBLP:conf/sc/UhlSHHKSS024}, including the principle of \emph{you only pay for what you use}.
It consists of composable pipe adapters inspired by \Cpp{} ranges that attach MPI metadata to arbitrary buffers, and thin wrappers around the core communication calls that enable operation-specific information exchange between send and receive sides.
Together these restore the convenience features of KaMPIng --- automatic receive count deduction, buffer resizing --- and, via move semantics and a move-only non-blocking handle, its memory safety guarantees.

\myintropar{Ecosystem Adapters (\cref{sec:ecosystem}).}
We demonstrate the extensibility of the core interface through first-class support for \emph{heterogeneous environments}.
Self-contained adapters for Kokkos, Thrust, and SYCL encapsulate the required interoperability logic and enable their direct use in arbitrary MPI operations, without modifying the core interface.

\begin{figure}[t]
\begin{lstlisting}[style=kamping]
namespace views = kamping::v2::views;

// (1) Core: STL containers work directly, no pointer extraction required
std::vector<double> v = ...;
mpi::send(v, dest, tag, comm);

// (2) KaMPIng-v2: alltoallv with automatic count exchange and buffer resizing
std::vector<int> local = ..., counts = ..., rbuf;
kamping::v2::alltoallv(local | views::with_counts(counts),
    rbuf | views::auto_counts | views::auto_displs | views::resize_v, comm);

// (3) Ecosystem: Kokkos views as first-class MPI buffers via core interface
Kokkos::View<double*> kv("data", n);
mpi::send(kv | views::kokkos, dest, tag, comm);
\end{lstlisting}
\caption{Representative examples spanning the three layers.
(1)~The core interface accepts STL containers directly.
(2)~\kampingtwo{} adds convenience: automatically derived displacements, receive count exchange and buffer resizing.
(3)~An ecosystem adapter makes a Kokkos view a first-class buffer for the core interface.}
\label{fig:teaser}
\end{figure}

\section{Related Work}\label{sec:related-work}
After the removal of the official MPI \Cpp{} bindings in MPI 3.0, numerous third-party libraries emerged to fill this gap.

Boost.MPI~\cite{Gregor2007} was the first wrapper supporting automatic type inference via Boost.Serialize and \lstinline{std::vector} overloads, but has not been maintained since 2008 and only covers MPI 1.1.
B-MPI3~\cite{Correa2018} is a \Cpp{}17 successor focused on iterator-based ranges, handling non-contiguous types via copying or Boost.Serialization.

MPP~\cite{DBLP:conf/pdp/PellegriniPF12} is a \Cpp{}11 library notable for its trait-based customization point: users specialize a trait template for their buffer types, and the library extracts the pointer, count, and datatype accordingly.
Although unmaintained since 2013 and limited to point-to-point communication, the approach offers a well-defined non-intrusive customization point.
We adopt, refine, and generalize this idea via \lstinline{buffer_traits<T>} and \lstinline{ptr()}, \lstinline{count()}, and \lstinline{type()} dispatch functions (see \cref{sec:data-buffer}).

RWTH-MPI~\cite{Demiralp2023} supports contiguous STL containers with automatic count and displacement inference for certain operations and PFR-based datatype generation.
It covers MPI 4.0, but its bindings closely mirror the C interface with little additional abstraction.

MPL~\cite{Bauke2015} is a header-only library based on a \emph{layout} abstraction that constructs views over contiguous memory regions expressible as MPI datatypes.
This suits regular communication patterns well but leads to verbose, potentially inefficient code for irregular communication~\cite{ghosh2021towards,beni2023empi}.

EMPI~\cite{beni2023empi} builds directly on OpenMPI rather than the C interface, eliding repeated runtime checks via unchecked lower-level primitives.
Communication context such as datatype, count, and tag are grouped into a \emph{message group} through which operations are dispatched.

KaMPIng~\cite{DBLP:conf/sc/UhlSHHKSS024} uses move semantics and a return-by-value design for a more ergonomic, memory-safe interface, including memory-safe non-blocking calls.
It supports compile-time datatype mapping for builtin types, optional PFR-based type derivation, and Cereal serialization for non-contiguous types.
Its named-parameter interface enables automatic count and displacement inference while preserving full low-level control.

KokkosComm~\cite{KokkosComm10740805} adds inter-node communication support for \lstinline{Kokkos::View} within the Kokkos performance portability framework~\cite{CarterEdwards20143202}, covering multiple backends including MPI and NCCL.

\section{A Standardizable Core for MPI C++ Bindings}\label{sec:core}
While the MPI standard describes its communication model in terms of typed data buffers, opaque object handles, and communication procedures, the C interface expresses these purely through raw pointers, integers, and opaque types, discarding structure the standard itself prescribes~\cite[Sec. 3.2.2]{mpi50}.
\Cpp{} allows us to stay closer to the actual \emph{semantic concepts} by modeling them at the interface level as \emph{\Cpp{}20 concepts}, building on and refining the conceptual groundwork of Avans et~al.~\cite{DBLP:conf/pvm/AvansCGSSSSU25}.
We introduce a concrete \emph{\Cpp{} core interface} modeled around buffers and objects that stays true to the MPI standard's semantics, but maps them to \Cpp{}.
More concretely, this means that every object that exposes memory, element count, and datatype through \lstinline{mpi::ptr()}, \lstinline{mpi::count()}, and \lstinline{mpi::type()} \emph{is} a data buffer.
This holds automatically for all \Cpp{} standard library types that users expect to work with MPI, and users can extend it to their own types through a dispatch mechanism customizable at compile time.
The resulting interface integrates seamlessly with existing \Cpp{} and MPI code and serves as our proposal for a future standard \Cpp{} MPI interface.

\subsection{A Semantic Model for MPI Data Buffers}\label{sec:data-buffer}

When using custom or standard library containers such as \texttt{std::vector<T>}, the MPI C interface forces user to extract raw pointers and size information manually, although the container already comprises the whole data context.
This is both non-ergonomic and conflicts with \Cpp{} Core Guidelines \cite[I.13]{CppCoreGuidelines}.
Most existing MPI \Cpp{} wrappers therefore add container support as their main abstraction layer~\cite{Bauke2015,Correa2018,Demiralp2023,Gregor2007,DBLP:conf/sc/UhlSHHKSS024}. 
However, this usually comes with significant shortcomings which are not compatible with a general MPI \Cpp{} interface:
They are either fixed to a limited set of container types, or decide to introduce hidden overheads by falling back to expensive serialization, derived data type construction or hidden allocations%
~\cite{DBLP:conf/pvm/AvansCGSSSSU25}.
These design decisions, motivated by specific use cases rather than general interface design, limit the applicability of these interfaces.

To solve this problem in a more systematic way, we identified, in previous work~\cite{DBLP:conf/pvm/AvansCGSSSSU25}, key rules and categories that qualify a \Cpp{} type for use as an MPI data buffer and proposed a corresponding set of \emph{\Cpp{} concepts} for modeling these constraints.
There, we showed that a type \lstinline{D} that is a \emph{contiguous} and \emph{sized} range, and where \lstinline{D::value_type} is either an MPI built-in type or the function \lstinline{mpi::type()} is defined for it, is a data buffer compatible with the MPI standard definition and can be used directly with MPI.
While this already covers the most common cases, such as STL container support, it has two shortcomings.
First, requiring types to implement the range interface is an unnecessary constraint:
types that hold contiguous data should be usable with MPI without being forced to expose iterator endpoints.
This rules out custom types without iterators, GPU containers such as \lstinline{thrust::device_vector} whose device pointers cannot be dereferenced on the host\footnote{While \texttt{thrust::device\_ptr<T>} can be dereferenced in host space (triggering an expensive single-element \texttt{cudaMemcpy}), it lacks support for \texttt{std::to\_address}, which is required by the contiguous range concepts, despite the underlying data being laid out consecutively in device memory.}, and special constructs like \lstinline{MPI_BOTTOM} that rely on absolute addressing.
Second, even the contiguous range abstraction is semantically wrong for MPI's full buffer model: MPI allows non-contiguous memory layouts via custom datatypes with offsets and gaps, as well as non-trivial displacements in collective operations with varying counts, which a contiguous range cannot express.

Therefore, we generalize our definition of data buffers, as shown in \cref{fig:data-buffer-concepts-v2}.
It follows the MPI standard's definition and the concept \lstinline{mpi::data_buffer} is satisfied if the functions \lstinline{mpi::ptr()}, \lstinline{mpi::count()}, and \lstinline{mpi::type()} are defined for a type.
We distinguish \lstinline{mpi::send_buffer} and \lstinline{mpi::recv_buffer} by constraining the address returned by \lstinline{mpi::ptr()} as being convertible to \lstinline{const void*} or \lstinline{void*}, respectively.

The key design element is the three-level dispatch behind \lstinline{mpi::ptr()}, \lstinline{mpi::count()}, and \lstinline{mpi::type()}, shown for \lstinline{mpi::ptr()} in \cref{fig:core_data_dispatch}.
Each function checks, in order: (1)~if a \lstinline{buffer_traits<T>} specialization for non-intrusive customization of types the user does not own exists; (2)~a member function such as \lstinline{mpi_ptr()} for types the user controls exists; or (3)~if the type is an MPI-compatibly STL range.
For contiguous ranges, \lstinline{mpi::ptr()} is defined via \lstinline{std::ranges::data()}, sized ranges default \lstinline{mpi::count()} to \lstinline{std::ranges::size()}, and
for ranges whose \lstinline{range_value_t} is an MPI builtin type, \lstinline{mpi::type()} automatically returns the matching \lstinline{MPI_Datatype}.
The generalized data buffer concept resolves both shortcomings: the trait level allows any type (GPU containers, \lstinline{MPI_BOTTOM}, arbitrary third-party objects) to become a data buffer without satisfying the range interface, while a \lstinline{buffer_traits<T>::type()} specialization returning a derived \lstinline{MPI_Datatype} covers non-contiguous memory layouts.
Since the fallbacks fire independently, a \lstinline{buffer_traits} specialization need only cover the accessors that cannot be derived automatically; for Thrust device vectors, for instance, only \lstinline{mpi::ptr()} requires a specialization since \lstinline{mpi::count()} and \lstinline{mpi::type()} remain satisfied by the range interface (see \cref{sec:gpu}).

\begin{figure}[t]
  \centering
  \begin{subfigure}[t]{0.50\linewidth}
\begin{lstlisting}[
style=kamping,
emph={[3]mpi,ptr,type,count,counts,displs}
]
template <typename T>
concept data_buffer =
 requires(T&& t, T const& ct) {
 { mpi::ptr(t) } -> /* ptr type*/;
 { mpi::count(ct) } -> /* int-like */;
 { mpi::type(ct) } -> /* conv. to */
                      /* MPI_Datatype*/;
};

template<typename T>
concept cnt_range = /* contig. range */
                    /* of int */;

template <typename T>
concept data_buffer_v =
 requires(T&& t, T const& ct) {
 { mpi::ptr(t) } -> /* ptr type*/;
 { mpi::type(ct) } -> /* conv. to */
                      /* MPI_Datatype*/;
 { mpi::counts(ct) } -> cnt_range;
 { mpi::displs(ct) } -> cnt_range;
};
  \end{lstlisting}
    \caption{Buffer concepts.
  }
  \label{fig:data-buffer-concepts-v2}
  \end{subfigure}%
  \hfill
  \begin{subfigure}[t]{0.46\linewidth}
\begin{lstlisting}[
style=kamping,
emph={[3]mpi,ptr,mpi},
emph={[2]mpi_ptr,std,ranges,data,buffer_traits},
escapechar=!
]
// (1) buffer_traits<T>
// specialization
template <typename T>
requires /* buffer_traits<T> */
auto mpi::ptr(T&& t) {
 return buffer_traits<T>::!\textcolor{my-teal}{ptr}!(t);
}

// (2) .mpi_ptr() member
// function
template <typename T>
requires /* .mpi_ptr() member */
auto mpi::ptr(T&& t) {
 return t.mpi_ptr();
}

// (3) contiguous range
// fallback
template <typename T>
 requires !\textcolor{black}{std}!::!\textcolor{black}{ranges}!::contiguous_range<T>
auto mpi::ptr(T&& t) {
 return std::ranges::data(t);
}
  \end{lstlisting}
  \caption{Three-level dispatch of \lstinline{mpi::ptr()}.}
  \label{fig:core_data_dispatch}
  \end{subfigure}
  \caption{The core buffer protocol.}
  \label{fig:buffer-core}
\end{figure}

\subsubsection{Data Buffers with Per-Process Counts and Displacements.}
\label{sec:data-buffer-v}
For collective operations with varying element counts per process, e.g., \lstinline{MPI_Gatherv} or \lstinline{MPI_Alltoallv}, MPI additionally requires per-process counts and displacements.
Since these are tightly coupled with the type and memory region, instead of following the MPI C interface and providing them as separate parameters, we extend \lstinline{data_buffer} to \lstinline{data_buffer_v}, as shown in \cref{fig:data-buffer-concepts-v2}.
This also facilitates working with custom buffer types that may already convey this information, which users would otherwise have to provide manually.

As with \lstinline{mpi::ptr()}, the functions \lstinline{mpi::counts()} and \lstinline{mpi::displs()} follow the same three-tiered dispatch and can be customized via \lstinline{buffer_traits<T>}.
Unlike plain \lstinline{data_buffer}, a scalar \lstinline{mpi::count()} is not required, and standard ranges do not satisfy \lstinline{data_buffer_v} automatically since they carry no per-process count or displacement information.
A valid \lstinline{data_buffer_v} can be constructed using \lstinline{mpi::mpi_span_v}, modeled after \lstinline{std::span}, but holding additional counts and displacements.

In \cref{sec:kamping-v2} we will show additional helpers provided by \kampingtwo{} to simplify the construction of buffers with attached counts and displacements.

\subsection{A Flexible Model for MPI Objects and Handles}\label{sec:objects}
The MPI standard defines a set of \emph{MPI objects}, such as communicators, data types, and requests.
In the C API, these are represented as \emph{opaque handles} that hide the implementation details from the user.
These handles are inexpensive to copy and compare, but must be explicitly created and destroyed using MPI calls.
Early works on C++ interfaces advocate for representing MPI objects as first-class C++ types that act as proxies for the underlying implementation-defined objects~\cite{Skjellum2001}. This approach has also been adopted by many application-specific MPI interfaces.

A native C++ object model further enables automatic resource management through RAII (Resource Acquisition Is Initialization)~\cite[16.5]{Stroustrup1994}\cite[E.6]{CppCoreGuidelines} and move semantics.
In this model, objects are constructed via \lstinline{MPI_*_create} in constructors and released via \lstinline{MPI_*_free} in destructors.
Ownership can be safely transferred through move operations, preventing resource leaks.

However, a purely RAII-based design is not feasible in practice due to MPI's reliance on global state.
In the world model, MPI provides global communicators such as \lstinline{MPI_COMM_WORLD} and \lstinline{MPI_COMM_SELF}, which are implicitly initialized and must not be freed.
Consequently, users cannot assume ownership of all MPI objects.
Furthermore, strict ownership models are problematic in library contexts, where object lifetimes are not always under the library's control.

To address these constraints, we adopt a more flexible design.
Using \lstinline{MPI_Comm} as representative example (the approach generalizes to other MPI object types), we provide both non-owning and owning abstractions:
\lstinline{mpi::comm_view} and \lstinline{mpi::comm}.
Both types expose common operations such as \lstinline{size()} and \lstinline{rank()} as member functions.
The non-owning \lstinline{mpi::comm_view} can be constructed from native MPI handles and is inexpensive to copy, as it does not impose lifetime semantics.
In contrast, \lstinline{mpi::comm} models ownership: it is move-only, manages resource lifetimes via RAII, and must either be constructed via library facilities or explicitly using \lstinline{mpi::comm::from_native()}.
Duplication is explicit, and owned objects can be converted to their corresponding view types.
Additionally, ownership can be relinquished explicitly via \lstinline{mpi::comm::release()}, which returns the underlying MPI handle and transfers responsibility for its lifetime to the caller, leaving the object in an empty state.
This separation allows users to either leverage automatic lifetime management or operate with lightweight, non-owning views, depending on their use case.

To ensure interoperability with existing code bases, we do not require users to adopt the object types provided by the core library.
In practice, many applications already define their own MPI wrapper types, and rewriting them to conform to a new interface would be impractical.
Instead, our interface is designed to operate seamlessly on user-defined types.
This is achieved by introducing the concepts \lstinline{mpi::convertible_to_mpi_handle<HandleType>} and \lstinline{mpi::convertible_to_mpi_handle_ptr<HandleType>}, which abstract over types that can be converted to a given native MPI handle type.
The conversion mechanism reuses the three-level dispatch strategy introduced for data buffers in \cref{sec:data-buffer}, while passing through native MPI handles unchanged.
This behavior, illustrated in \cref{fig:object-dispatch}, enables seamless integration of user-defined types and existing MPI code.

Overall, the design avoids enforcing a single representation or ownership model, instead providing interoperable abstractions that integrate with both native MPI and user-defined types.

\begin{figure}
  \centering
  \begin{subfigure}[b]{.48\linewidth}
\begin{lstlisting}[
style=kamping,
emph={[3]comm_view,comm},
emph={[2]cv, c},
escapechar=!
]
std::vector<int> sdata = ...;

// (1) native MPI handle (baseline)
mpi::send(sdata, dst,
          tag, !\textcolor{my-teal}{MPI\_COMM\_WORLD}!);

// (2) non-owning view
// (no lifetime semantics)
mpi::comm_view cv {MPI_COMM_WORLD}; 
mpi::send(sdata, dest, tag, cv);

// (3) owning RAII object
// (lifetime-managed)
mpi::comm c = cv.dup();
mpi::send(sdata, dest, tag, c);
\end{lstlisting}
  \end{subfigure}
  \begin{subfigure}[b]{.48\linewidth}
\begin{lstlisting}[style=kamping,showlines=true,
emph={[3]handle_traits,handle},
emph={[2]my_comm},
]
// (4) user-defined communicator
// (non-intrusive adaptation)
template<>
struct mpi::handle_traits<MyComm> {
 static MPI_Comm handle(
  MyComm const& comm) {
   return comm.mpi_communicator();
 }
};

MyComm my_comm = ...;
mpi::send(sdata, dest,
          tag, my_comm);

\end{lstlisting}
  \end{subfigure}
  \caption{Uniform use of MPI communicators across native handles, non-owning views, owning RAII objects, and user-defined types.}
  \label{fig:object-dispatch}
\end{figure}

\subsection{The Core Interface}\label{core:interface}

Based on the abstractions for memory and objects proposed in the previous sections, we introduce the MPI C++ core library interface.
This layer is designed as a standardizable C++ binding to MPI that preserves interoperability with existing MPI and C++ codebases, without redefining the MPI semantics.
Instead, it provides a thin C++ layer that directly reflects the abstractions defined by the MPI standard.
These abstractions are extended with sensible defaults for C++ standard library types and idiomatic C++ patterns.

All MPI procedures operate on arbitrary data buffers satisfying the \lstinline{send_buffer} or \lstinline{recv_buffer} concepts (and their \lstinline{_v} variants) introduced in \cref{sec:data-buffer}, replacing the explicit (pointer, count, datatype) tuple with a uniform buffer abstraction, enabling automatic standard container support.
All arguments representing MPI objects are modeled via the concept \lstinline{convertible_to_mpi_handle<HT>} introduced in \cref{sec:objects}, enabling uniform use of native MPI handles, core library owning and non-owning object types, and user-defined wrapper types.
Auxiliary parameters such as ranks and tags accept any type convertible to an integer; tags additionally support enumeration types for improved type safety. Additional customization points allow user-defined rank and tag representations.
This ensures that auxiliary MPI semantics remain fully extensible while preserving simple integer-based usage as the default case.
Error handling follows the MPI configuration: if \lstinline{MPI_ERRORS_ARE_FATAL} is set, errors terminate the program; if \lstinline{MPI_ERRORS_RETURN} is enabled, errors are mapped to C++ exceptions. This enables idiomatic return-by-value interfaces for object operations such as \lstinline{mpi::comm::rank()} or \lstinline{mpi::comm::split()} while preserving access to runtime errors.
Since destructors of owning object types must not throw, failures during calls to \lstinline{MPI_*_free} under \lstinline{MPI_ERRORS_RETURN} result in program termination. For scenarios requiring explicit recovery (e.g., fault tolerance), owning objects provide a \lstinline{.free()} function that releases the underlying MPI resource and leaves the object in an empty state.
This mapping ensures that the core interface remains a thin and predictable C++ projection of the MPI execution model, rather than introducing an independent error-handling abstraction layer.

\Cref{fig:core-example} shows the full implementation of \lstinline{mpi::allgatherv()} and demonstrates how the concept-based design enables a direct and type-safe mapping of MPI C calls into C++.
All functionality is implemented uniformly in terms of the dispatch functions like \lstinline{mpi::ptr}, \lstinline{mpi::count}, \lstinline{mpi::type}, and \lstinline{mpi::handle}, which form a small, uniform set of customization points with well-defined default behavior.
This makes the dispatch system the central architectural mechanism of the core interface.
Together, these design choices define a minimal and stable interface that is sufficiently expressive for direct use, while remaining fully aligned with the MPI standard.
It intentionally omits higher-level concerns such as runtime inference, automatic buffer resizing, and ownership management for non-blocking operations --- not as limitations, but as a deliberate separation of concerns.
\kampingtwo{}, introduced in \cref{sec:kamping-v2}, builds on this foundation to recover the ergonomics and safety guarantees of KaMPIng~\cite{DBLP:conf/sc/UhlSHHKSS024}, demonstrating that the core interface is a sufficient and stable base for richer C++ abstractions.

\begin{figure}
  \centering
\begin{lstlisting}[style=kamping,
escapechar=\#,
emph={[2],send_buffer,recv_buffer_v,convertible_to_mpi_handle,},
emph={[3]mpi,count,ptr,type,counts,displs,handle}
]
template <
  #\textcolor{my-teal}{mpi}#::send_buffer                         SBuf,
  #\textcolor{my-teal}{mpi}#::recv_buffer_v                       RBuf,
  #\textcolor{my-teal}{mpi}#::convertible_to_mpi_handle<MPI_Comm> Comm>
void allgatherv(SBuf&& sbuf, RBuf&& rbuf, Comm const& comm) {
    int err = MPI_Allgatherv(mpi::ptr(sbuf), mpi::count(sbuf), mpi::type(sbuf),
        mpi::ptr(rbuf),
        std::ranges::data(mpi::counts(rbuf)),
        std::ranges::data(mpi::displs(rbuf)),
        mpi::type(rbuf),
        mpi::handle(comm)
    );
    if (err != MPI_SUCCESS) {
        throw #\textcolor{black}{mpi}#::mpi_error{err};
    }
}
\end{lstlisting}
  \caption{Concept-based implementation of \lstinline|mpi::allgatherv|, illustrating the mapping of MPI C calls to the C++ core interface using the dispatch system.}
  \label{fig:core-example}
\end{figure}

\section{KaMPIng-v2: A Composable and Ergonomic MPI Interface}\label{sec:kamping-v2}

The KaMPIng library~\cite{DBLP:conf/sc/UhlSHHKSS024} has been an important step towards modern \Cpp{} language bindings for MPI and offers an ergonomic, memory-safe MPI interface via named-parameter emulation with automatic type inference, count and displacement deduction, and memory-safe non-blocking calls.
These features enable writing concise, high-level MPI code while preserving low-level control. 
Its named-parameter design, however, relies on complex template meta-programming and shifts the interface away from MPI's own notion of a data buffer as a coherent (pointer, count, datatype) triple.

We present \kampingtwo{}, a redesign of KaMPIng that retains its safety and usability goals while aligning closely with the core interface introduced in \cref{sec:core}.
It consists of three components: (1)~pipe adapters in the style of \lstinline{std::views} that attach or override MPI metadata (datatype, count, per-rank counts and displacements), turning arbitrary objects into data buffers or modifying existing ones, and composing freely with standard library views (\cref{sec:adapters}).
Additionally, it provides thin wrappers around the core interface's MPI procedures, adding (2)~operation-specific metadata exchange between send and receive sides, enabling automatic buffer sizing and count exchange (\cref{sec:deferred}), and (3)~opt-in memory-safety support for non-blocking calls (\cref{sec:nonblocking}).
\kampingtwo{} also demonstrates that the core interface is a sufficient and stable base for richer \Cpp{} abstractions.

\subsection{Buffer Adapters}\label{sec:adapters}
The \Cpp{}20 standard library's ranges library~\cite{niebler2018ranges} introduced \emph{range adapters}: lazy, composable transformations over ranges expressed via pipe syntax.
An adapter such as \lstinline{std::views::take(n)} does not copy or modify data, but produces a lightweight view over the first \lstinline{n} elements of a range, which can then be passed directly wherever a range is expected.
Adapters are also composable: \lstinline{vec | std::views::drop(2) | std::views::take(n)} produces a view over \lstinline{n} elements of \lstinline{vec} starting at offset 2.

Since the core library's buffer concepts (see \cref{sec:data-buffer}) are satisfied by any contiguous sized range with a builtin element type, standard library views that preserve the contiguity property (such as \lstinline{std::views::take}, \lstinline{std::views::drop} or \lstinline{std::span}) can be passed directly to core MPI C++ interface introduced above, as demonstrated in \cref{fig:adapters} (1).

Standard library adapters operate on the range structure but do not explicitly carry the MPI-specific metadata modeled by the \lstinline{mpi::data_buffer} and \lstinline{mpi::data_buffer_v} concepts.
Therefore, \kampingtwo{} provides a complementary set of adapters that attach or override MPI properties.
They allow modifying element count, datatype, per-rank counts and displacements while preserving the same pipe syntax and composing freely with standard library views (see \cref{fig:adapters}, (2)--(4)).
All \kampingtwo{} adaptors produce types that fully satisfy the core buffer concepts and require no changes to the core interface.
This mechanism in turn lifts MPI buffer construction to a composable, declarative style analogous to modern C++ range pipelines.

More concretely, each adapter is a lightweight view that wraps its input and passes through all buffer accessor functions (\lstinline{mpi::ptr()}, \lstinline{mpi::count()}, \lstinline{mpi::type()}, etc.) unchanged, overriding only the specific accessor it is responsible for.
For example, \lstinline{views::with_type(dt)} adds an \lstinline{mpi_type()} member returning \lstinline{dt} while delegating all other accessors to the underlying buffer unchanged.
Adapters that compute derived properties do so lazily:
\lstinline{views::auto_displs()} computes displacements as the exclusive prefix sum of the counts only when queried, and only once.
This mirrors the laziness guarantee of \lstinline{std::ranges} views and ensures that adapters in a pipeline incur no overhead unless their output is actually consumed.

A recurring challenge in MPI is lifecycle management of derived datatypes:
custom \lstinline{MPI_Datatype} handles must be explicitly committed before use and freed afterwards.
Building on the type pool concept introduced in~\cite{DBLP:conf/pvm/AvansCGSSSSU25}, \kampingtwo{} provides \lstinline{type_pool}, a move-only registry that owns committed \lstinline{MPI_Datatype} handles for its lifetime, providing RAII semantics for MPI datatype management.
The pipe adaptor \lstinline{views::with_auto_pool(pool)} (see \cref{fig:adapters} (5)) lazily registers and attaches the correct \lstinline{MPI_Datatype} for the buffer's value type; \lstinline{with_pool} is available for cases requiring explicit pre-registration.
Using the pool requires users to specialize \lstinline{mpi::types::mpi_type_traits} to describe the type's memory layout -- either manually or automatically via \lstinline{struct_type}, which derives the layout from \lstinline{T}'s fields using Boost.PFR~\cite{Polukhin2016} without any manual field specification, with native \Cpp{}26 reflection~\cite{childers2025reflection} as a future replacement.

\begin{figure}[t]
\begin{lstlisting}[
style=kamping,
emph={[3]std,vector,take,type_pool,mpi_type_traits,types,views, with_type, auto_displs, with_counts, with_auto_pool}
]
namespace views = kamping::v2::views;
std::vector<double> data = ...; std::vector<int> counts = ...;

// (1) std::views::take - send first n elements
mpi::send(data | std::views::take(n), dest, tag, comm);

// (2) with_type - override MPI datatype
mpi::send(data | views::with_type(MPI_DOUBLE), dest, tag, comm);

// (3) Compose STL and kamping views: send n elements with custom type
mpi::send(data | std::views::take(n) | views::with_type(my_type),
          dest, tag, comm);

// (4) Variadic recv buffer: attach counts, auto-compute displacements
mpi::allgatherv(data,
 data | views::with_counts(count) | views::auto_displs(), comm);

// (5) type_pool: automatic MPI datatype lifecycle management.
// Specialize mpi_type_traits using struct_type - Boost.PFR reflects
// the fields of MyStruct automatically:
struct MyStruct { int x; double y; };
template <>
struct mpi::types::mpi_type_traits<MyStruct>
  : public mpi::types::struct_type<MyStruct> {};

kamping::v2::type_pool pool;
std::vector<MyStruct> v = ...;
// MPI_Datatype for MyStruct committed on first use,
// freed when pool is destroyed
mpi::send(v | views::with_auto_pool(pool));
\end{lstlisting}
  \caption{Standard library and \kampingtwo{} buffer adaptors. Examples (1) uses standard library views that satisfy
   the core buffer concepts directly. Examples (2)--(3) attach MPI-specific metadata via \kampingtwo{} adaptors. Example
   (4) constructs a variadic receive buffer for \lstinline{allgatherv} using composed adaptors.
   Example (5) demonstrates automatic MPI datatype management via \lstinline{type_pool}.
 }\label{fig:adapters}
\end{figure}

\subsection{Ergonomic Receive Buffers}\label{sec:deferred}
The adapters described above are self-contained: each derives its output solely from the buffer it wraps, without any communication.
In practice, users often prefer not to supply all metadata explicitly: receive sizes can be derived from the incoming message, and per-rank counts in collectives can be gathered from the sending side.
\kampingtwo{} addresses this through an \lstinline{infer()} customization point, called by each \lstinline{kamping::v2::} wrapper before calling into the underlying core interface MPI operation, with access to all send and receive buffers and the communicator.
Its behavior can be specialized per operation and buffer type.

By default, it supports working with \emph{deferred buffers}, which do not now their receive count(s) in advance.
They expose a protocol to set the count from the outside, which is called from \lstinline{infer()}.
Depending on the operation, \lstinline{infer()} resolves the missing information without any additional communication (e.g.\ for \lstinline{allreduce}, the received count equals the send count), via \lstinline{MPI_Probe} for point-to-point receives, or through a preceding count exchange for collectives such as \lstinline{allgatherv}.

Consistent with the laziness of the adapter pipeline, \lstinline{infer()} only records the resolved count in the view; the actual \lstinline{resize(n)} on the underlying container is triggered lazily by \lstinline{mpi_ptr()}, immediately before MPI writes into the buffer.
The mechanism follows the \emph{you only pay for what you use} principle: \lstinline{infer()} is a no-op for plain buffers, so callers that supply all metadata explicitly pay no overhead.

\kampingtwo{} also recovers the Cereal serialization support of the original KaMPIng implementation as a composable view (\lstinline{views::serialize}) rather than logic baked into the communication layer --- enabled by the deferred buffer and data buffer concepts, it is available uniformly across all MPI operations.

\subsection{Safe Non-blocking Operations}\label{sec:nonblocking}

Non-blocking MPI operations initiate communication and return a request handle without completing it.
MPI semantics require that buffers involved in an in-progress operation remain unmodified until the request completes, but this constraint cannot be enforced at the language level without library support.
This idea was first introduced in KaMPIng~\cite{DBLP:conf/sc/UhlSHHKSS024} and formalized in~\cite{DBLP:conf/pvm/AvansCGSSSSU25}; \kampingtwo{} realizes it on top of the buffer concept and view pipeline introduced above.

We address memory-safety for non-blocking communication through perfect forwarding of buffer arguments combined with a move-only \lstinline{iresult} handle.
When a buffer is passed as an rvalue, it is moved into \lstinline{iresult}, which owns it for the duration of the operation; when passed as an lvalue, it is borrowed and the caller retains ownership.
The buffer is only accessible via \lstinline{.wait()}, which completes the request and returns owned buffers by value, or \lstinline{.test()}, which returns \lstinline{std::optional} containing the buffer only if the request has already completed.
Invalid access to in-progress owned buffers is thus prevented through library semantics, as depicted in \cref{fig:non-blocking}.

\begin{figure}[t]
\begin{lstlisting}[style=kamping,
emph={[3]move,wait,irecv},
emph={[2]req}
]
// rvalue receive buffer is moved to the returned iresult ...
std::vector<int> rbuf(rcount);
auto req = kamping::v2::irecv(std::move(rbuf), source, tag, comm);
// ... and is only moved back to caller after calling .wait()
rbuf = req.wait();
\end{lstlisting}
\caption{Non-blocking \lstinline{irecv}:  a receive buffer is moved into \lstinline{iresult}, which prevents access until \lstinline{.wait()} returns it.}
  \label{fig:non-blocking}
\end{figure}

\section{Bridging to Performance Portability and GPU Ecosystems}\label{sec:ecosystem}

GPU computing and performance portability have become central concerns in the world of HPC.
Many MPI implementations support GPU-aware communication, transferring data directly between device memory regions without routing through the host.
However, the MPI interface still requires raw pointers and explicit size and type information, forcing users to extract metadata manually from GPU containers, undermining the abstractions these libraries provide.

The data buffer concepts introduced in \cref{sec:data-buffer} solve this cleanly:
adapting a GPU container to the buffer protocol is sufficient to make it a first-class citizen in MPI, and the resulting adapter works uniformly across all MPI operations without any changes to the core interface.
In the simplest case, no adapter is needed at all --- pointers to device memory allocated via \lstinline{cudaMalloc} or \lstinline{sycl::malloc_device} can be wrapped in \lstinline{std::span} and passed directly to any MPI call.
In the following we demonstrate more complex adapters for different programming models:
For Kokkos~\cite{CarterEdwards20143202}, a widely-used performance portability framework, we provide a composable adapter for \lstinline{Kokkos::View} that builds on KokkosComm~\cite{KokkosComm10740805} to handle both contiguous and non-contiguous views, the latter via packing into a contiguous buffer.
For CUDA and SYCL, we show that \lstinline{thrust::device_vector} requires only a two-line \lstinline{buffer_traits} specialization, while \lstinline{sycl::buffer} admits an adapter that naturally aligns with SYCL's accessor model.

\subsection{Kokkos}\label{sec:kokkos}
KokkosComm~\cite{KokkosComm10740805} addresses the challenges of combining Kokkos and MPI by accepting \lstinline{Kokkos::View} directly in communication calls, handling non-contiguous views via transparent packing.
However, its design is operation-centric: each MPI operation must be explicitly implemented with Kokkos support.

We take a different approach: rather than wrapping individual operations, we wrap the view itself into \lstinline{views::kokkos}, a type that satisfies the buffer concepts and therefore works with all MPI operations automatically.
The packing strategy follows KokkosComm: contiguous views are passed directly after fencing the execution space; non-contiguous views are lazily packed into a contiguous scratch buffer via \lstinline{Kokkos::deep_copy} on first \lstinline{mpi_ptr()} access and unpacked after the call.
Since \lstinline{kokkos_view} satisfies the buffer concepts, it works directly with the core interface as shown in \cref{fig:ecosystem-kokkos}, giving Kokkos support across all MPI operations without any changes to the core --- including operations KokkosComm does not yet support.
The view type composes naturally with the existing \kampingtwo{} adapters, allowing users to augment it with custom types, counts, or displacements.
Using the \kampingtwo{} communication interface further enriches this: \lstinline{kokkos_view} supports resizing the underlying view, making it composable with \lstinline{| views::resize} just like any standard container.
Since buffer ownership and lifetime are handled transparently in the \kampingtwo{} interface layer, non-blocking operations work with Kokkos views without any additional modification, matching the lifetime guarantees KokkosComm provides.

\begin{figure}[t]
\centering
  \begin{subfigure}[t]{0.48\linewidth}
\begin{lstlisting}[style=kamping,showlines=true,
emph={[3]views,kokkos,resize},
emph={[2]Kokkos,View}
]
namespace views = kamping::v2::views;
Kokkos::View<int*> sv(n);
mpi::send(sv | views::kokkos,
          dest, tag, comm);
Kokkos::View<int*> rv(0);
kamping::v2::recv(
 rv | views::kokkos | views::resize,
 src, tag, comm);

  \end{lstlisting}
  \caption{Kokkos.}
  \label{fig:ecosystem-kokkos}
  \end{subfigure}
\begin{subfigure}[t]{0.48\linewidth}
\begin{lstlisting}[style=kamping,
escapechar=!,
emph={[3]views},
emph={[2]sycl,buffer,accessor}
]
sycl::buffer<int> buf(...);
q.submit([&](!\textcolor{black}{sycl}!::handler& h) {
 sycl::accessor acc{buf, h};
 h.host_task(
   [=](!\textcolor{black}{sycl}!::interop_handle ih) {
     mpi::send(acc | views::!\textcolor{my-purple}{sycl}!(ih),
               src, tag, comm);
 });
});
\end{lstlisting}

  \caption{SYCL.}
  \label{fig:ecosystem-sycl}

\end{subfigure}
\caption{Ecosystem adapters: Kokkos (left) and SYCL (right).}
\label{fig:ecosystem}
\end{figure}

\subsection{GPU Memory}\label{sec:gpu}
For raw CUDA memory allocated via \lstinline{cudaMalloc}, we already discussed that it can be supported by our interface by wrapping it in a \lstinline{std::span}.
The Thrust library~\cite{bell2012thrust,thrust}, a CUDA-based parallel STL-like library by NVIDIA, offers an abstraction for managing device memory via \lstinline{thrust::device_vector}.

Thrust device vectors do not model the \lstinline{mpi::data_buffer} concept directly, since \lstinline{.data()} returns a \lstinline{thrust::device_ptr} rather than a raw pointer.
This can be mitigated by adding a \lstinline{mpi::buffer_traits} specialization that extracts the underlying raw pointer. Defining \lstinline{mpi::buffer_traits::ptr} is sufficient, all other accessor functions are satisfied automatically by \lstinline{thrust::device_vector}'s sized-range interface.
With that, Thrust works with any MPI operation through GPU-aware MPI, composing naturally with the \kampingtwo{} adapter library.

SYCL~\cite{sycl2020} is a heterogeneous programming model for CPUs and GPUs based on standard \Cpp{}, providing a single-source, task-based abstraction for execution across devices from different vendors.
It supports two main memory management approaches: traditional pointer-based device allocations and \lstinline{sycl::buffer}, a high-level representation of data which is not necessarily tied to a single memory location.
Data represented by buffers can only be accessed via accessor objects, which ensure that the data is available when used.
Consequently, MPI calls operating on buffer-managed data residing on a device need be enqueued as \emph{host tasks}, which execute on the host and expose backend-native pointers via a \lstinline{sycl::interop_handle}.
Our adapter for \lstinline{sycl::buffer} aligns with this model:
inside a host task, \lstinline{acc | views::sycl(ih)} constructs a buffer exposing the native pointer through the interop handle, as shown in \cref{fig:ecosystem-sycl}.
This respects SYCL's memory safety guarantees, as the pointer is accessed only within the host task scope in which it is valid.

\section{Conclusion and Future Work}\label{sec:conclusion}
We presented a layered \Cpp{} MPI interface built on \Cpp{}20 concepts.
The core layer formalizes the MPI standard's own buffer abstraction as the \lstinline{data_buffer} concept, backed by a three-level dispatch that covers STL containers automatically and admits non-intrusive customization for third-party types.
A parallel abstraction for MPI object handles enables uniform use of native handles, RAII owning objects, and user-defined wrapper types.

\kampingtwo{} demonstrates that this core is a sufficient and stable base for richer abstractions: composable pipe adapters bring the ergonomics of \lstinline{std::ranges} to buffer construction; a deferred-buffer protocol enables automatic count deduction and buffer resizing following the \emph{you only pay for what you use} principle; move semantics support memory safety for non-blocking calls.
Adapters for Kokkos, Thrust, and SYCL confirm the core's extensibility: a minimal \lstinline{buffer_traits} specialization or a self-contained adapter is sufficient to make GPU containers first-class MPI buffers across all operations, without modifying the core interface.

The intentionally minimal design of the core layer lends itself to standardization and opens the door for third-party ecosystem adapters building on the abstractions.
Our current proposal focuses on two-sided communication, but the concepts naturally translate to one-sided MPI\@.
We will address improved \Cpp{} semantics for that in the future.
Moreover, other language bindings like \texttt{rsmpi}~\cite{Steinbusch2015} and \texttt{MPI.jl}~\cite{Byrne2021} implicitly define similar trait-based buffer protocols, indicating cross-language convergence.
Formalizing and aligning these across languages is a natural direction for future work.
Additionally, lifting the buffer abstraction to the interface level preserves type and layout information that is erased in the MPI C interface, supporting compiler-assisted program analysis and verification tools that currently require ad-hoc workarounds.

\begin{credits}
  \subsubsection{\ackname}
  The authors gratefully acknowledge the Gauss Centre for Supercomputing e.V. (\url{www.gauss-centre.eu}) for funding this project by providing computing time on the GCS Supercomputer SuperMUC-NG at Leibniz Supercomputing Centre (\url{www.lrz.de}).
  
  The authors gratefully acknowledge the computing time provided on the high-performance computer HoreKa by the National High-Performance Computing Center at KIT (NHR@KIT). This center is jointly supported by the Federal Ministry of Education and Research and the Ministry of Science, Research and the Arts of Baden-Württemberg, as part of the National High-Performance Computing (NHR) joint funding program (\url{https://www.nhr-verein.de/en/our-partners}). HoreKa is partly funded by the German Research Foundation (DFG).

  The authors thank all participants of the MPI Forum's Language Binding Working Group for valuable feedback and discussions.
  
\subsubsection{\discintname}
  The authors have no competing interests to declare that are relevant to the content of this article.
\end{credits}
\bibliographystyle{splncs04}
\bibliography{kampingv2.bib}
\end{document}

%% file: preamble.tex
\usepackage{graphicx}
\usepackage{subcaption}

\usepackage{csquotes}
\usepackage[capitalise]{cleveref}

\usepackage[textsize=scriptsize]{todonotes}
\todostyle{nu}{color=red,backgroundcolor=red!20,bordercolor=red,author=NU}
\todostyle{ms}{color=blue,backgroundcolor=blue!20,bordercolor=blue,author=MS}
\todostyle{}{color=green,backgroundcolor=green!20,bordercolor=green,author=DB}

\definecolor{my-dark-red}{RGB}{183, 28, 28}
\definecolor{my-red}{RGB}{244,67,54}
\definecolor{my-pink}{RGB}{233,30,99}
\definecolor{my-purple}{RGB}{156,39,176}
\definecolor{my-deep-purple}{RGB}{103,58,183}
\definecolor{my-indigo}{RGB}{63,81,181}
\definecolor{my-blue}{RGB}{33,150,243}
\definecolor{my-light-blue}{RGB}{3,169,244}
\definecolor{my-cyan}{RGB}{0,188,212}
\definecolor{my-teal}{RGB}{0,150,136}
\definecolor{my-green}{RGB}{76,175,80}
\definecolor{my-light-green}{RGB}{139,195,74}
\definecolor{my-lime}{RGB}{205,220,57}
\definecolor{my-yellow}{RGB}{255,235,59}
\definecolor{my-amber}{RGB}{255,193,7}
\definecolor{my-orange}{RGB}{255,152,0}
\definecolor{my-deep-orange}{RGB}{255,87,34}
\definecolor{my-brown}{RGB}{121,85,72}
\definecolor{my-grey}{RGB}{158,158,158}
\definecolor{my-blue-grey}{RGB}{96,125,139}
\definecolor{my-lipics-grey}{rgb}{0.6,0.6,0.61}
\usepackage{url}
\usepackage{listings}
\usepackage[scaled=0.85]{beramono}
\lstdefinestyle{kamping}{
basicstyle={\ttfamily\scriptsize},
emphstyle={\bfseries\color{my-teal}},
emphstyle={[2]\color{my-teal}},
emphstyle={[3]\color{my-purple}},
emphstyle={[4]\color{my-indigo}},
keywordstyle={\bfseries\color{blue}},
commentstyle=\textcolor{gray},
morekeywords={constexpr, size_t, concept, requires, template, typename},
stringstyle={\color{my-green}},
}

\newcommand{\Cpp}{C++}
\newcommand{\kampingtwo}{KaMPIng-v2}

\usepackage{lmodern}
\usepackage[%
activate={true,nocompatibility},%
final,%
tracking=true,%
kerning=true,%
spacing=true,%
stretch=10,%
shrink=30]{microtype}